# Securities Lending Haircuts and Indemnification Pricing

Wujiang Lou[1]

August 28, 2020; Updated September 2021

**Abstract**

Securities borrowing and lending are critical to proper functioning of securities markets. To alleviate securities owners' exposure to borrower default risk, overcollateralization and indemnification are provided by the borrower and the lending agent respectively. Haircuts as the level of overcollateralization and the cost of indemnification are naturally interrelated: the higher haircut is, the lower cost shall become. This article presents a method of quantifying their relationship. Borrower dependent haircuts satisfying the lender's credit risk appetite are computed for US Treasuries and main equities by applying a repo haircut model to bilateral securities lending transactions. Indemnification is designed to fulfill a triple-A risk appetite when the transaction haircut fails to deliver. The cost of indemnification consists of a risk charge, a capital charge, and a funding charge, each corresponding to the expected loss, the economic capital, and the redundant fund needed to arrive at the triple-A haircut.

**1. Introduction**

An arbitrager or hedger wishing to take a short position in a security needs to borrow that security in order to sell it in the market. Market makers, no longer paid to carry deep inventories nowadays, borrow securities as well, even if customer flows are merely lightly imbalanced. Being able to borrow securities is critical to the smooth functioning of the securities market (Duffie, Garleanu, and Pedersen, 2002). Foley-Fisher, Gissler, and Verani (2019) show evidence that AIG (American International Group)'s near collapse in 2008 and subsequent suspension of its massive securities lending operations severely impacted liquidity of the US corporate bond market. Typically, the borrowing activity involves a (beneficiary) securities owner, a borrower, and an agent who facilitates securities lending and borrowing transactions. The short sellers, however, are typically not a party to the transaction: their brokers or prime brokers are. On the lender side, customers of brokerage houses can opt to participate in their lending programs, but the securities lending market is dominated by mutual funds, pension funds and endowments, investment funds, banks and insurance companies, and government entities (Baklanova, Caglio, Keane, and Porter, 2016), who employ securities lending as a return enhancement investment strategy.

---

[1] The views and opinions expressed herein are the views and opinions of the author, and do not reflect those of his employer and any of its affiliates.

Securities lenders are exposed to borrower default risk: borrower not returning the security upon request or an outright default when the loaned securities are not called back. The lender would have to redeem its cash reinvestment if cash collateralized or to liquidate non-cash collateral securities, in order to purchase replacement securities in the open market. It is possible that the market has moved adversely and the collateral proceed is short of the repurchase price. The lender will then suffer a default loss.

To mitigate the risk of loss, borrowers are required to post up collateral in excess of the loaned securities market value. For example, borrowing $100 worth of US Treasury bonds would require $102 of cash collateral. The ratio of the collateral value to the securities value is termed margin ratio or simply margin[2]. The term *haircut* is often used interchangeably with margin, although most will drop the 100%, i.e., 102% margin or 2% haircut. In that sense, haircut is excess margin. Circumstances could arise when these haircut levels are not sufficient to avoid losses. In the industry's common practice, lending agents offer borrower default indemnification: any loss arising from the over-collateralized securities loan upon a borrower default will be paid by the agent.

According to a US government pilot data collection summarized by Baklanova *et al* (2016), the seven major lending agents report that, on year end 2015, 97.16%[3] of agency lending activities employ indemnification. Indemnification of course does not come for free. Considering that the surveyed lending dollar volume already exceeds 1 trillion and that agents not included in the pilot may not be less likely to use indemnification, indemnification pricing is a critical task in securities lending. As an integral part of the securities lending compensation[4], high indemnification cost will impede the securities lending market and the functioning of the securities market. To reduce the cost, the lending agents have strengthened all aspects of credit risk management, including borrower credit analysis, borrower selection, collateral criteria, and securities concentration limit setting. Ultimately though, the cost of indemnification needs to be quantified.

---

[2] In the standardized legal document for securities lending, the Master Securities Lending Agreement (MSLA, SIFMA, 2000), the official term used is market value percentage.
[3] Percentage computed using Table 11 and Table 3 of Baklanova *et al*, 2016.

Horner (2013) shows that the value of indemnification could be a small fraction of the regulatory cost of a bank offering indemnification in the post-crisis capital regime. The value is the expected loss of the lending, with the borrower default probability computed over the period of the 5 day buy-in period and its exposure computed as a call option on the securities stroked at the margin and termed for the buy-in period. A reasonable value for stock lending is estimated at 0.2 basis points (bps), while the regulatory capital charge estimated under BASEL III could be 10.3 bps, a significant portion of the existing lending fee structure. The cost of indemnification obviously will depend on the margin ratio: a higher margin should result in lower indemnification cost. The exact mechanism has not been studied in the literature. In fact, there is no formal publications on the subject of borrower default indemnification valuation and/or pricing, except for Horner (2013)'s research piece.

This paper contributes to the field by presenting a new method of pricing securities lending indemnification. Recognizing that both haircuts and indemnification are means of credit enhancement and that owners naturally expect from indemnification an `AAA` rating on the lending transaction, we first seek a haircut ceiling that would have resulted in an `AAA` rated credit risk profile without indemnification. We then set up a hypothetic independent account to fund the gap between the transaction haircut and the haircut ceiling. The cost of indemnification is thus transformed to the cost of setting up and servicing the account, which consists of a risk charge, a capital charge, and a funding charge. The risk charge is a running spread reflecting the expected loss of the lending transaction, the capital charge is a cost of capital computed as the expected shortfall of the transaction, and the funding charge compensates the use of fund for the redundant reserve part of the account.

---

[4] When the collateral is cash, the lender invests it in some short-term cash-equivalent instruments to earn interest and agrees to pay the borrower a rebate, at a rate lower than the earned rate. The difference between the earned rate and the rebate rate is the lending compensation. If the collateral is non-cash, the borrower agrees to pay the lender a loan fee as compensation, quoted as a rate based on the daily market value of the loaned securities. Agents and owners will have fee split agreements that include the cost of indemnification.

## 2. Rating Targeted Securities Lending Haircut

Since we are speaking of borrower default risk, haircut as a measure of credit enhancement is conceptually borrower dependent, as well as securities dependent. In the agent lending market, however, it has been practically a fixture. Typical margins are 102% for domestic securities and 105% for foreign securities (Lavender, 2011, and Blackrock, 2020). But, in retail brokerage, individual investors are required to maintain a minimum margin of 125%, with an initial margin of 150%. While this is rules-based margin (or Reg T margin) suitable for retail investors, hedge funds as the largest group on the short sales side are subject to risk-based margin, which is determined by their prime brokers, often using a value-at-risk (VaR) based measure in combination with their ratings. Because hedge funds are unrated, banks act as intermediaries: borrowing securities in the agency lending market, then lending to hedge funds. Needless to say, the full spectrum of securities borrowers also includes publicly rated financial firms who borrow for their own accounts.

A securities lending haircut model, therefore, will need to take different borrower credit into account. It is possible that haircut's sensitivity with respect to borrower credit quality might vary, depending on loaned securities and other considerations. In some circumstances, sensitivities may turn out to be marginal or negligible, and haircuts are better fixed for operational convenience. Lou (2020) develops a repo haircut model that addresses the relationship between counterparty credit and repo haircuts, which we can extend to securities lending transactions.

In the course of a securities lending transaction, both the borrower and the lender may default, but it is customary to only consider the borrower default and to treat the lender as a going concern. In the event of default, the transaction is immediately terminated. All accrued and unpaid fees are due immediately, e.g., the loan fee payable to the lender and the rebate fee to the borrower.

If the borrower is unable to transfer the loaned securities back to the lender[5], the lender can choose to buy back the securities in the open market and net the purchase money against the collateral it holds. If the collateral is non-cash, it can also dispose it in the market. This buy-in or default settlement process, however, could take days to complete. If the market price goes up beyond the margin, the lender suffers a loss, which can be claimed through to the defaulted borrower's estate.

Similar to repo's margin period of risk (MPR), a securities lender's exposure window or the buy-in period is longer than one day. MPR, a critical input to the model, is the time period starting from the last date when margin is met to the date when the defaulting counterparty is closed out with completion of asset and collateral disposal. MPR could cover a number of events or processes, including valuation, margin calculation, margin call, default notification and default grace period, and finally time to buy back. The exact length of MPR depends on legal and process controls. BASEL III has established a minimum MPR of 5 business days for securities financing transactions[6], but it is understood not applicable to securities lending transactions with cash collateral. For US Treasuries and US equities, given the depth of the markets, we will use 3-day MPR.

For margining purposes, the security is fair valued, or valued at the mid-price. Upon borrower default, the lender has to buy back at the ask side of the price. We use $g$, $1 > g \geq 0$, to denote this ask to fair spread. Obviously, $g$ links to the liquidity of the security, i.e., the less liquid a security is, the greater $g$ is. In what is called a specific wrong way risk situation, closing out a default counterparty borrowing a large proportion of an issued security may cause a price surge. This can be captured by a jump-on-default, which in the simplest form can be incorporated into $g$.

To summarize, we consider a borrower default time at $\tau$, when the margin is last met, an MPR of $u$ during which there is no margin posting, and finally the securities are bought back at time $\tau + u$ instantaneously at the market with a possible premium $g$. If we denote the security market value as *B(t)*, the cash collateral value the defaulting counterparty (borrower) posts as of time *t* is *E(t)*, *E(t)=(1+h)B(t)*, where *h* denotes the constant haircut, *1>h ≥0*. The margin agreement is assumed to have a zero minimum transfer amount. The lender's residual exposure is *(B(τ+u)(1+g) - E(τ))⁺*. Denote *R* borrower's recovery rate, *Γ(t)* default indicator, 1 if *τ≤t* , 0 otherwise, we write the loss function at time *t* as follows,

$$L(t) = (1 - R)\Gamma(t)\big(B(\tau + u)(1 + g) - B(\tau)(1 + h)\big)^+ \tag{1}$$

---

[5] Sec lending has gone beyond the traditional short selling use case where a defaulted short seller couldn't buy back the security and transfer it back to the lender. Nowadays, securities can be borrowed and used as collateral in the borrower's other transactions, e.g., OTC derivatives. In such a case, collateral return is possible if proper segregation is in place.

[6] In the banking regulatory space, securities lending transactions with cash collateral are not required to establish minimum haircut floors, subject to certain cash reinvestment standards and liquidity/maturity matches, while non-central bank repos are subject to regulatory minimum haircut floors. This is a consideration given to its critical role in market functioning rather than a consideration of its risk (definitely not suggesting that banks lend securities to hedge funds at zero haircut).

If we wish not to consider the borrower's credit support, we can set *t=0*, *Γ(t)=1*, *R=0*, and the loss function is purely driven by the securities' return over a period of MPR,

$$L(u) = (1 + g)B_0 \left(\frac{B(u)}{B_0} - \frac{1+h}{1+g}\right)^+ \tag{2}$$

where *B₀=B(0)*. Comparing to repo's one period loss function: $L(u) = (1 - g)B_0 \left(\frac{1-h}{1-g} - \frac{B(u)}{B_0}\right)^+$, two differences are easily noted. First, we add *g* and *h* in securities lending, but subtract them in repo. In fact, *g* and *h* can take algebraic form to apply to both. Second, the repo loss is on the down side of the return, while the securities lending loss is on the up side. Therefore, the repo haircut model can be rather trivially extended to securities lending.

Once we have determined the loss distribution, haircuts can be designed to achieve predetermined wholesale credit rating quality. For example, haircuts can be defined on meeting expected loss (EL) target *L₀*,

$$h_{EL} = \inf\{h \in R : E[L|h] \leq L_0\}, \tag{3}$$

*L₀* can be set based on EL criterion of certain designated high credit rating, whether internal or external, e.g., Moody's Aa1. With the first dollar loss probability (or probability of default) criterion, adopted by rating agencies such as Standard and Poor's (S&P), one would want to control *Prob(L>0)*. Lou (2020) has a third criterion which uses economic capital (EC) as the target measure.

The loss function is governed by a joint asset and credit dynamic model. The double exponential jump-diffusion (DEJD, Kou, 2002) model is used for asset dynamics and the Black-Karasinski or log OU (Ornstein-Uhlenbeck) process is used for credit spread, as follows,

$$X(t) = \log\left(\frac{B(t)}{B_0}\right) = \mu t + \sigma_a(\rho W(t) + \sqrt{1-\rho^2}W_a) + \sum_{j=1}^{N(t)} Yj, \tag{4}$$

$$dlog(\lambda(t)) = k(\bar{y} - log(\lambda(t)))dt + \sigma dW(t) \tag{5}$$

where σ$_a$ is the asset volatility, μ the return of the asset, $\rho$ correlation between credit and the securities price, $k$ the mean reversion rate of the default intensity λ(t), $\bar{y}$ the mean reversion level, $\sigma$ credit spread volatility. $W_a(t)$ and $W(t)$ are two independent Brownian motions. $N(t)$ is a Poisson process with intensity $\lambda_a$, $Y_j$ is a random variable denoting the magnitude of the $j$-th jump. $Yj, j=1, 2, ...$, are a sequence of independent and identically distributed mixed-exponential random variables with the pdf $f_Y(x)$ given by

$$f_Y(x) = p_u \eta e^{-\eta x} I\{x \geq 0\} + q_d \theta e^{\theta x} I\{x < 0\} \tag{6}$$

where $p_u$ and $q_d$ are up jump and down jump switching probabilities, $p_u+q_d=1$. The up-jump mixture exponentially is distributed at a rate of $\eta>1$. Similarly, $\theta>0$ is the down jump mixture's rate. DEJD's asymmetric up and down jump specification is a perfect fit as repo's exposure is on the down side (thus a put option) while sec lender's exposure is on the up side (a call option).

The model has a sophistication level that requires quite extensive numerical computations. The original Monte Carlo simulation procedure to seek unconditional loss distribution and expected loss is replaced by a Karhunen-Loeve decomposition scheme (Lou, 2021) that transforms the procedure to a numerical integration using Gauss-Hermite quadrature. Computational time is saved dramatically and most computations can be done in a few seconds.

## 3. Indemnification Pricing

Securities owners are generally investors not equipped well enough to conduct advanced credit analysis. Most remain comfortable with and rely on public or private credit ratings. For investors seeking borrower default indemnification, the usual thinking in terms of credit quality is that of `AAA` (triple-A). Accordingly, we set the rating criterion to be that of `AAA`. The EL targets taken from Moody's idealized ELs for 1 year time horizon are 3.00E-07, 3.10E-06, 7.50E-06, and 1.66E-05 for 'Aaa', 'Aa1', 'Aa2', and 'Aa3' respectively (Bielecki 2008). Denote the triple-A haircut as $\dddot{h}_c$,

$$\ddot{h}_c = \inf\{h \in R: E[L_c|h] \leq 3 * 10^{-7}\} \tag{7}$$

where subscript letter `c` denotes the credit grade of the borrower. Obviously, (7) can be easily changed to use PD based rating criterion, such as S&P's ratings.

From the investor's perspective, if securities are loaned to a triple-A rated borrower, he is fine with zero haircut. If the borrower's credit rating is lesser, additional margin would be required to make the overall lending credit risk profile meet the triple-A criterion. Conceptually, as the borrower's credit deteriorates, the amount of haircut would increase. The lending haircut model can be used to create a schedule of $\ddot{h}_c$, per counterparty credit grade.

Let $h_c$ be the agreed/transacted lending haircut. If $h_c \geq \ddot{h}_c$, the investor doesn't need indemnification, as the overall credit risk profile of the transaction is already at the level of triple-A. If $h_c < \ddot{h}_c$, the triple-A minded investor would ideally want to (re)negotiate with the borrower to raise the transaction haircut $h_c$. From the investor's perspective, he is indifferent whether the borrower or the agent makes up the haircut gap of $\ddot{h}_c - h_c$. Therefore, the purpose of an indemnity can be seen as a credit enhancement to bring up the whole credit risk profile back to that of triple-A. Consequentially, the cost of indemnification is the cost of posting and maintaining the gap amount.

We can think of this gap amount being set in an independent account. The gap amount is decomposed into three components: expected loss, economic capital (EC) per expected shortfall (ES) calculation, and the remaining balance ($= \ddot{h}_c - h_c - EL - ES$) called the redundant fund as it is nearly risk-free after EL and ES have been taken out. The haircut model computes lending EL and EC/ES given a haircut $h_c$. A lending agent can thus proceed to price indemnification from this perspective. EL leads to a pass-through charge or risk charge that needs to be collected back from the policy buyer. ES forms a capital usage charge usually at the cost of capital, e.g., 15%. The last component is credit risk-free yet totally illiquid, thus subject to a liquidity (funding) charge from the indemnification provider as cash needs to be set aside. The fair policy rate *S* of indemnification becomes the sum of risk charge, capital charge at cost pf capital $s_c$, and funding charge at the rate of $s_f$,

$$S = EL + ES * s_c + (\ddot{h}_c - h_c - EL - ES) * s_f \tag{8}$$

The lending agent should demand a greater compensation than this rate from the borrowers and subtract it out when negotiating the fee split with the owners. If the lending fee is not sufficient, the agent would need to negotiate harder for higher haircuts, which would reduce the policy rate. Therefore, this can play an indirect role in setting securities lending transaction haircuts.

Our approach is different from Horner (2013) in a number of prospects: $1^{st}$, our asset price dynamics is a jump diffusion process rather than the standard geometric Brownian motion. $2^{nd}$, we don't assume independent default probability and asset price. $3^{rd}$, we compute and price in the impact of economic capital rather than regulatory capital. Economic capital is a universal risk measure, but BASEL 3 regulatory capital is a bank supervisory measure, thus not applicable to non-bank lending agents. Economically, EC arises because, like repos, securities lending transactions are neither hedgeable nor diversifiable, so that a capital measure has to be employed and compensation for use of capital needs to be priced in. Lastly, our model integrates haircuts and indemnification as two different means of credit enhancement, as it is based on an underlying securities lending haircut model, extended from the repo haircut model of Lou (2020).

The haircut model is a fairly complex model. From the implementation point of view, its technical aspect should be manageable as its components are standard models and its performance has been improved by developing a fast numerical procedure for the joint loss distribution (Lou 2021) to replace Monte Carlo simulation. Inputs of the model include the following: six asset price parameters (equations 4 and 6), three borrower CDS curve parameters (equation 5), one correlation between credit and the asset price, and MPR. The double exponential jump diffusion model is estimated historically by the maximum likelihood estimation approach with likelihood functions numerically computed via the two-sided Laplacian inversion of the pdf or via the explicit formulae obtained by Ramezani and Zeng (2007).

CDS parameters can be estimated from corporate bonds or CDS data of the borrower. This is not a problem in the agency lending market when the borrowers are large broker-dealers whose credit products are traded in the credit market. For borrowers without actively traded CDS, e.g., banks' primer brokerage clients, large lenders or lending agents are typically capable of conducting credit risk analysis to come up with appropriate internal ratings. CDS or credit spread proxies can be obtained by pivoting on internal credit ratings among firms with and without credit spreads curves. For certainly class of borrowers, e.g., 'B' rated or lower, it might be prudent not to consider their credit strength and compute haircuts on a borrower independent basis.

Another practical concern has something to do with the traditional business view and practice that securities financing transactions are considered simple, not sophisticated enough to call for advanced modeling efforts and analytics. The landscape has changed quite a bit since the financial crisis of 2008-2009, as regulators have been extremely concerned with the instability of short-term funding and securities lending markets, which triggered the crisis. A model like this can help the industry evolve and adapt to the post-crisis regime where securities financing transactions are treated increasingly in line with OTC derivatives. The model provides ways to examine and quantify the risk of securities lending and enhances our capability for better and more efficient risk management. For instance, facing competition from an old-timer who quotes the standard 102% equity margin on a high volatility stock which the model calls for 108%, a trader may still transact at that level, but can structure the trade by reducing trade size, mingling with other low volatility stocks, demanding intra-day margin, and charging higher fee to compensate the increased tail risk (EC). Even if none of these are possible, the trader would at least know where he's taking up the risk and can stay on extra due diligence.

## 4. Computational Results

To satisfy FSB (2015)'s haircuts requirements of a 5-year history consisting of at least one stress period, for computations presented herein, we choose the 5-year period from 1/2/2008 to 1/2/2013 when the financial market has significant stress in the second half of 2008 and early 2009.

### 4.1. Triple-A lending haircuts without borrower credit support

To study the impact of the buy-in period (MPR), we compute securities lending haircuts assuming MPRs of 1 day, 3 days, and 5 days. Target credit ratings are triple-A, with both PD (aka S&P) and EL (aka Moody's) rating criteria. To focus on MPR, we consider the extreme case where the borrower is assumed credit worthless. For given MPRs, these haircuts are the maximum required to arrive at investors' triple-A expectation.

Table 1. US main equities securities lending haircuts with MPR=1 day, 3 days, and 5 days targeting triple-A rating in PD and EL based rating methodologies respectively.

|    | 1 day | 3 days | 5 days |
|----|-------|--------|--------|
| PD | 7.15  | 11.96  | 15.72  |
| EL | 8.61  | 13.59  | 17.53  |

Table 1 shows the most challenging part of any haircut model, parametric or non-parametric: MPR as a critical input is neither a market observable nor a transaction attribute. Unfortunately, it has to come from an agent or a bank's comprehensive assessment covering legal language, default settlement procedure, loaned securities liquidity including size and concentration, and past experiences. Figure 1 compares repo haircuts and securities lending haircuts that don't utilize counterparty credit support. Their differences are small enough, although sec lending haircuts tend to be slightly greater than repo haircuts.

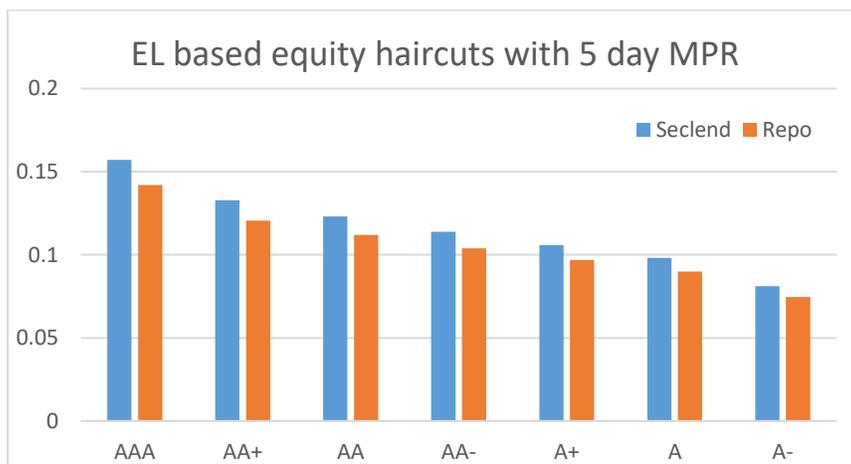

Figure 1. Comparisons of predicted US main equities securities lending haircuts and repo haircuts when targeting top rating band of 'A3' to 'Aaa'.

## 4.2. Bilateral securities lending haircuts

For bilateral securities lending transactions, the lender can use the model to seek desired haircut levels based on internal wholesale lending guidance. Firms, for instance, could standardize wholesale lending at certain rating level, e.g., `A`. Shown in Figure 2 are bilateral haircuts facing a roughly `BBB` rated borrower with 5-year CDS priced at 250 bps. The effect of the counterparty credit support is about 4% haircut saving, when compared to counterparty independent haircuts also shown in the Figure. The impact of correlation shown as the comparison of 80% vs zero correlation is moderate, for the loss exposure lies in the tail of asset returns where the Gaussian component of the asset dynamics is not expected to matter. The general wrong way risk therefore is limited, obviously due to the short-term effect and the much less pronounced loss after truncation with haircuts.

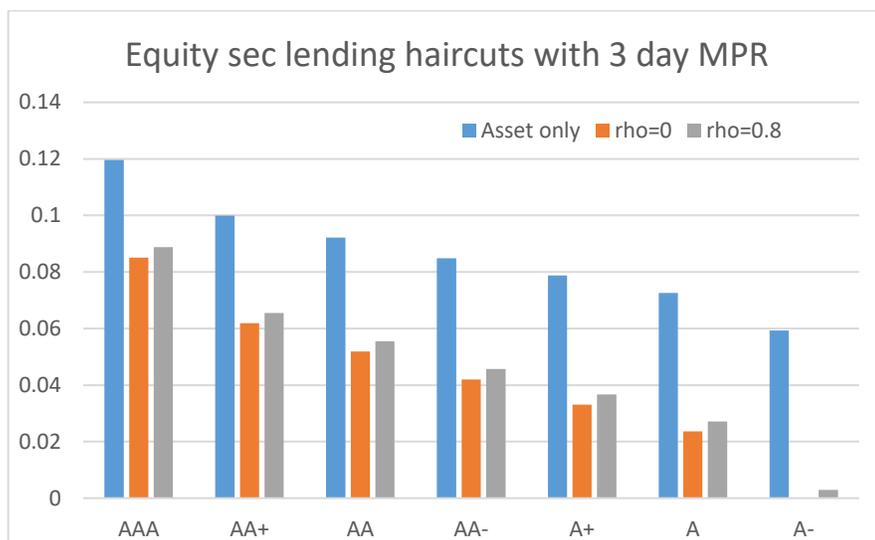

Figure 2. Securities lending haircuts to a `BBB` rated borrower with CDS 250 bp. Counterparty independent haircuts are shown under the label of 'Asset only'.

Baklanova *et al* (2016) report a wide range of equity securities lending margins, with a mean of 103%, 5% percentile at 102% and 95% percentile at 110%. The mean of 3% lines up about `A+' target rating at 3-day MPR. The reported 95% percentile haircut of 110%, however, is not comparable as it does not distinguish cash loans from non-cash collateral loans[7], while our results are cash loan haircuts. Out of the 1,018.6 billion 3-month average total market value of securities on loan during Q4 2015, 869.1 billion are borrowed by registered broker-dealers and another 142 billion by banks and credit unions. Hence, it is reasonable for us to assume the borrower has an average rating of `BBB`.

---

[7] As the authors point out, variations of margins around the mean can be attributed to the comingling of cash and non-cash loans. For non-cash loans, lending agents may have bundled the haircut on the non-cash collateral into the overall margin, e.g., 105% cash margin may be reported as 110% margin if the non-cash collateral securities have a haircut of 5%.

US Treasury securities are the most popular in the securities lending space (Baklanova *et al* 2016), followed by foreign government debts, and US equities. Other fixed income securities are also possible, including corporate bonds, municipal bonds, and mortgage or asset backed securities. Bond haircuts are seldom determined per bond, or per cusip using industry jargon, obviously because bonds have fixed maturity dates and a particular issuance has limited circulation in the secondary market. As a result, bond haircuts are broadly based on similar bonds which allow a bond price index to be used as the proxy. Since the exposure window is a short MPR, there is no need for yield dynamics, so long as the model could be separately calibrated or estimated from the most relevant historic data to reflect its duration bracket, e.g., historical prices of bonds with residual maturity of 5 years used to estimate the DEJD model for determination of haircuts of the same maturity bracket.

Table 2. US Treasury securities lending haircuts, with MPR=3 days, sample counterparty credit ratings from 'A' to 'D', and EL rating targets from 'AAA' to 'A-'.

| Cpty\Target | AAA  | AA+  | AA   | AA-  | A+   | A    | A-   |
|-------------|------|------|------|------|------|------|------|
| A           | 2.8  | 1.39 | 0.88 | 0.39 | 0    | 0    | 0    |
| BBB         | 3.3  | 1.85 | 1.33 | 0.87 | 0.46 | 0.04 | 0    |
| BB          | 3.85 | 2.37 | 1.83 | 1.36 | 0.98 | 0.61 | 0    |
| B           | 4.45 | 2.96 | 2.41 | 1.92 | 1.53 | 1.17 | 0.41 |
| D           | 5.97 | 4.47 | 3.91 | 3.4  | 2.98 | 2.59 | 1.81 |

Applicable UST haircuts are very small for investment grade borrowers ('A' and 'BBB' rated in the table). With an 'AA' target, for example, the securities lend haircut rounds to 1%. In practice, zero haircut is the rule when facing investment banks and large broker-dealers. For asset managers, especially hedge funds, 5% haircuts dominate, which agrees with the 'AAA' target level for borrowers rated 'B' or 'D'. Unfortunately, there is no data collection available that could cover these types of lower credit grade borrowers, so a thorough empirical study is desirable but unfeasible at this time.

For non-cash collateralized loans, a separate haircut to those collateral security could be determined using the repo haircut model. Because the collateral securities are usually subject to substitution in kind, it is difficult to price in precisely the correlation between the loaned securities and the collateral securities. Fortunately, the impact of correlation might be limited as shown in Figure 2, and computing haircuts separately could be prudent at least.

**4.3. Indemnification pricing example**

Below we conduct a pricing exercise (Table 3) where investors will accept S&P triple-A rated indemnification. The borrower is an average major broker-dealer, `BBB` rated. The loaned securities are US main equities and will transact at 5% haircut with cash collateral.

Table 3. Sample indemnification pricing sheet.

| | |
|---:|---:|
| Margin | 105% |
| S&P AAA | 8.1% |
| haircut gap | 3.1% |
| EL | 0.0009% |
| ES | 0.8946% |
| Funding | 2.2045% |
| cost of capital | 15% |
| funding cost | 1% |
| risk chrg (bps) | 0.09 |
| capital chrg (bps) | 13.42 |
| funding chrg (bps) | 2.20 |
| total (bps) | 15.72 |

We first use the haircut model to find the triple-A haircut facing the borrower at 8.1% under 3-day MPR. A haircut gap of 3.1% (=8.1%-5%) thus exists which needs to be capitalized and funded. At 5% haircut, the expected loss is calculated at 9.33e-6, or 0.0933 bps. ES is 0.8946%, which results in a capital charge of 13.42 bps at 15% cost of capital. The redundant fund amount is 2.2045% (=3.1%-0.8946%). Assuming 1% funding spread, this leads to 2.20 bps funding charge. Adding these three charges together arrives at a total indemnification charge of 15.72 bps. This pricing is dominated by the capital charge, which comes in the same ballpark with Horner (2013)'s estimate of regulatory capital charge under BASEL 3.

Obviously, such a pricing sheet relies on the critical input of the buy-in period. If the buy-in period is longer, say 5 days, the total charge will be 34.64 bps. Table 4 shows the final pricing for 2%, 3%, and 5% haircuts with `A' rated and `BBB` rated borrowers, using both S&P rating methodology and Moody's methodology. Assuming that the industry largely works on about 35 bps (Baklanova *et al*, 2016) lending compensation at 105% margin, 3-day MPR is more realistic than 5-day MPR, as it would leave about 20 bps of fee to be split between the agent and the owners. At lower margins, the indemnification cost is higher and the net lending fee gets squeezed to an extent that may no longer make sense to lend out securities. That being the case, Table 4 shows that the correct way to keep securities lending alive is to move up the margin rate: reducing indemnification charge by roughly 9 bps per 1% of margin increase.

Table 4. Securities lending indemnification pricing scenarios, using S&P and Moody's triple-A rating criteria, assuming `A` or `BBB` rated borrowers, under 3 or 5 day buy-in periods.

|  |  | 5 MPR | | | 3 MPR | | |
|---|---|---|---|---|---|---|---|
|  | Borrower | 2% | 3% | 5% | 2% | 3% | 5% |
| S&P | A | 52.8 | 43.1 | 24.0 | 35.8 | 26.2 | 8.4 |
|  | BBB | 64.0 | 54.1 | 34.6 | 44.7 | 34.9 | 15.7 |
| Moody's | A | 55.4 | 45.7 | 26.6 | 38.0 | 28.4 | 10.6 |
|  | BBB | 66.5 | 56.6 | 37.1 | 46.9 | 37.1 | 17.9 |

## 5. Concluding Remarks

We examine securities lending transactions' risk from the securities lender's or lending agent's perspective and find that the modeling technique introduced earlier for repos can be trivially extended to securities lending. The haircut model can be used for bilateral securities lending transaction pricing to produce borrower sensitive margins.

A common practice and requirement in the agent lending market is borrower default indemnification, which serves as the next in line (after haircut) defense against borrower default. We propose a novel pricing method that integrates these two means of credit enhancement: haircut and indemnification. Specifically, a lending agent can hypothetically capitalize and fund a margin gap, which, if added to the transaction margin, would render a triple-A rated whole risk profile to the satisfaction of the securities owners. The cost of indemnification thus consists of a risk charge associated with the expected lending loss, a capital charge related to the economic capital reserve, and a funding charge for the remaining risk-free redundant fund.

Sample computations of bilateral securities lending haircuts and agent indemnification are given in the limited context of the available market survey. A full-scale empirical study of the model would be desired.

**References**

Baklanova, V., C. Caglio, F. Keane, and B. Porter (2016), A Pilot Survey of Agent Securities Lending Activity, FRBNY Research Report, August 2016.

Bielecki, T. (2008), Rating SME transactions, Moody's Investor Services, May 2008.

Blackrock (2020), Securities Lending Viewed through the Sustainability Lens, Policy Spotlight, February 2020.

Duffie, D., N. Garleanu, and L. H. Pedersen (2002), Securities lending, shorting, and pricing, Journal of Financial Economics, Vol 66 (2002), pp 307-339.

Foley-Fisher, N., S. Gissler, and S. Verani (2019), Over-the-counter market liquidity and securities lending, Review of Economic Dynamics Vol 33 (2019), pp 272–294.

FSB (2015), Transforming shadow banking to resilient market-based finance: regulatory framework for haircuts on non-centrally cleared securities financing transactions, *Financial Stability Board*.

Horner, G. (2013), The Value and Cost of Borrower Default Indemnification, State Street's Digest of Topics in Securities Finance, November, 2013.

Kou, S.G. (2002), A Jump-Diffusion Model for Option Pricing, Management Science, 48 (8), pp 1086-1101.

Lavender, J.M. (2011), A Best Practice Oversight Approach for Securities Lending, J. P. Morgan Thought, Summer 2011, pp 28-31.

Lou, W. (2020), Repo Haircuts and Economic Capital: A Theory of Repo Pricing, Journal of Credit Risk forthcoming, https://papers.ssrn.com/sol3/papers.cfm?abstract_id=2725633, SSRN preprint.

Lou, W. (2021), Fast Computation of Securities Financing Loss Distribution in Joint Lognormal Credit and Jump Diffusion Asset Model (June 26, 2021). Available at SSRN: http://dx.doi.org/10.2139/ssrn.3874679.

Ramezani, C.A., and Y. Zeng (2007), Maximum likelihood estimation of the double exponential jump-diffusion process, Annals of Finance, Vol 3(4), pp 487-507.

SIFMA (2000), Master Securities Lending Agreement (MSLA), 2000 version. Securities Industry and Financial Markets Association (SIFMA), available at https://www.sifma.org/wp-content/uploads/2017/08/MSLA_Master-Securities-Loan-Agreement-2000-Version.pdf.